\documentclass[fleqn,twocolumn,amsmath,amssymb,longbibliography]{revtex4-1}

\usepackage{xifthen}
\newcommand{\refAppendix}[6]{#1
  \ifthenelse{\isempty{#2}}%
    {}
    {\protect\cite{#2}}
    #3\protect\ref{#4}#5#6\xspace
  }
  
\usepackage{forloop,ifthen} 

\usepackage{amsfonts}
\usepackage{mathtools}
\usepackage{amsmath}
\usepackage{amssymb}

\usepackage{graphicx}
\usepackage{times}
\usepackage[colorlinks=true,linkcolor=blue,citecolor=red,plainpages=false,pdfpagelabels]{hyperref}
\usepackage{xspace}

\usepackage{color}

\usepackage[normalem]{ulem}

\providecommand{\U}[1]{\protect\rule{.1in}{.1in}}

\usepackage{multirow} 

\begin{document}

\title{Experimental Demonstration of Topological Charge Protection in Wigner Current}

\author{Yi-Ru Chen,$^{1}$ Hsien-Yi Hsieh,$^{1}$ Jingyu Ning,$^{1}$ Hsun-Chung Wu,$^{1}$ Hua Li
  Chen,$^{2}$ You-Lin Chuang,$^{3}$ Popo Yang,$^{1}$ Ole Steuernagel,$^{1, 4}$ Chien-Ming Wu,$^{1}$ and
  Ray-Kuang Lee$^{1,2,3,5}$}

\affiliation{$^{1}$Institute of Photonics Technologies, National Tsing Hua University, Hsinchu 30013, Taiwan\\
$^{2}$Department of Physics, National Tsing Hua University, Hsinchu 30013, Taiwan\\
$^{3}$Physics Division, National Center for Theoretical Sciences, Taipei 10617, Taiwan\\
$^{4}$Department of Physics,~Astronomy~and~Mathematics,~University~of~Hertfordshire,~Hatfield, AL10 9AB, UK\\
$^{5}$Center for Quantum Technology, Hsinchu 30013, Taiwan}
 \email{rklee@ee.nthu.edu.tw}

\date{\today}
\begin{abstract}
  We experimentally reconstruct Wigner's current of quantum phase space dynamics for the first time.
  We reveal the ``push-and-pull" associated with damping and diffusion due to the coupling of a
  squeezed vacuum state to its environment. In contrast to classical dynamics, where (at zero
  temperature) dissipation only ``pulls" the system toward the origin of phase space, we also
  observe an outward ``push" because our system has to obey Heisenberg's uncertainty relations.
With squeezed vacuum states generated by an optical parametric oscillator at variable pumping
  levels, we identify the pure squeezing dynamics and its central stagnation point with a
  topological charge of `$-1$'. We experimentally verify that this charge is protected for weakly as well as strongly
  decohering conditions. This work demonstrates high resolving power and establishes an experimental paradigm for measuring
  quantumness and non-classicality of the dynamics of open quantum systems.
\end{abstract}
\maketitle

We establish experimentally that quantum dynamics in phase space can be
studied directly and in great detail, using Wigner's quantum phase space distribution~$W$ and the
associated phase space current~$\vec{J}$ which governs~$W$'s time evolution. We do this in a
quantum optical system using a degenerate squeezer setup.


Using Wigner's representation of quantum systems~\cite{Wigner_PR32}, based on Wigner's
distribution $W(x,p)$~\cite{Royer_PRA77} in quantum phase space (with coordinates~$x$ for position and~$p$ for
momentum), makes it easier to compare classically with quantum states than the commonly used density matrix $\rho$~\cite{Leonhardt_PQE95,Schleich_01,Zurek_NAT01,Zachos_book_05}.  The associated
Wigner current~$\vec{J}$, similarly, allows for a direct visualization of the system dynamics and
its comparison with classical Hamiltonian flows~\cite{Ole_PRL13,Kakofengitis_EPJP17}.  No such
current exists to describe $\rho$'s evolution.

Any experimental quantum mechanical system is unavoidably subject to a number of dissipative
processes which, for our example, degrades squeezing, resulting in anti-squeezing that is always larger than the
squeezing.  Losses and phase noise have to be taken into account to quantify this type of
degradation of quantum states~\cite{LIGO-quantum}.


With the help of machine learning, our neural network-enhanced tomo\-graphy has demonstrated good
performance when extracting detailed information about the degradation in a system undergoing
decoherence dynamics~\cite{QST-ML}.  Here, we go one step further by reconstructing not only the
Wigner distribution~$W$, but also the associated Wigner current~$\vec{J}$ experimentally.  In order
to access physically relevant structures in quantum phase space, our experimental reconstruction of
Wigner current allows identifying pure squeezing dynamics as well as dissipation and decoherence
currents.

\begin{figure}[t]
\centering
\includegraphics[width=8.4cm]{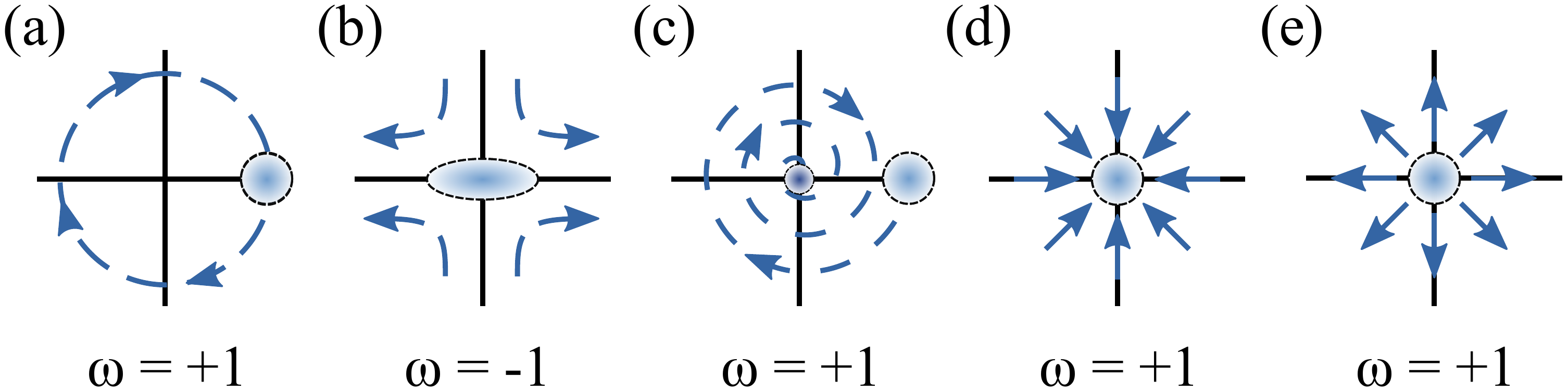}
\caption{Schematics of phase space dynamics and the corresponding Wigner currents for (a) a coherent state, (b) a squeezed vacuum state (in a co-rotating frame), (c) the classical picture of a damped coherent state, (d) the damping current (with inward `pull'), and (e) the diffusing current (with outward `push'). $\omega$ is the winding number~(\ref{eq:WindingNumber}) of $\vec{J}$'s stagnation point at the origin.}
\label{fig:Stagnation_Point_Current_Schematic}
\end{figure}

Naively, dissipation drives the system toward the origin in phase space~\cite{open} by classical
damping currents, as illustrated in Fig.~\ref{fig:Stagnation_Point_Current_Schematic}. But, Heisenberg’s uncertainty relation imposes a minimum
phase space area on the wave function, even for the (squeezed) vacuum state at zero temperature,
prohibiting such purely classical dynamics.
Here, we experimentally reveal the ``push-and-pull" associated with this damping and diffusion due to the
coupling of a squeezed vacuum state to its thermal environment. 

The interplay of the corresponding
contributions in the Wigner current~$\vec{J}$ allows us to display the phase space currents which,
in the steady state, give rise to detailed balance and the Einstein coefficients. This demonstrates that our approach provides a powerful diagnostic toolbox for probing
details of a quantum state's dynamics.\\

\noindent {\it Wigner current in squeezers.}\textemdash 
After three decades of conceptual innovations and technological improvements~\cite{Slusher, LAWu}
squeezed vacuum states~\cite{Yuen, SQ-book, SQ-30}, with up to $15$~dB squeezing,
have been demonstrated~\cite{15dB}.  
To go beyond the standard quantum limit in measurements, squeezed vacuum states have been widely
used in quantum metrology~\cite{metro-1, metro-Science, metro-NP, metro-NP2} and advanced
gravitational wave detectors ~\cite{Cave, GW-13, GW-18, GW-19, GW-Virgo, GW-FDSQ-1, GW-FDSQ-2}. 
Here, we exploit the excellent control of such squeezer systems, which allows us to experimentally determine Wigner's distribution~$W$ and reconstruct Wigner's phase space current~$\vec{J}=(J_x,J_p)$ and its effects on the evolution of the system, with high resolution. Together, $W$ and $\vec{J}$ fulfil the continuity equation~\cite{Oliva_PhysA17}
\begin{eqnarray}\label{eq:continuity}
\frac{\partial W(x,p)}{\partial \tau_{\text{eff}}} + \vec{\nabla} \cdot \vec{J}(x,p) = 0,
\end{eqnarray}
where, $\vec{\nabla} = (\partial_x, \partial_p)$ and $\tau_{\text{eff}}$ an effective time specified
below. Wigner's continuity equation~(\ref{eq:continuity}) is the equivalent of the conventional von
Neumann's equation (with Lindblad master equation terms)~\cite{Cabrera_PRA15} that describes the
quantum dynamics of the open system state with the density matrix $\rho(t)$
\begin{eqnarray}
\frac{d \rho}{dt}&=&-\frac{i}{\hbar}[H,\rho]+\frac{\gamma}{2}(\bar{n}+1)\left(2 a\rho a^{\dagger}-a^{\dagger}a\rho-\rho a^{\dagger}a\right)\cr
&&+\frac{\gamma}{2}\bar{n}\left(2a^{\dagger}\rho a-a a^{\dagger}\rho-\rho a a^{\dagger}\right).
\label{mastereq}
\end{eqnarray}

Unlike this conventional approach, using Eq.~(\ref{eq:continuity}) has the great advantage of
allowing us to visualize quantum dynamics in phase space since $W$ and $\vec{J}$ are real-valued and
exist everywhere. Indeed, the corresponding line integrals along $\vec{J}$ yield field lines,
reminiscent of classical phase portraits~\cite{Ole_PRL13,Kakofengitis_EPJP17}.

Here we study squeezed vacuum states which can be produced by systems predomi\-nant\-ly generating
photon pairs, e.g., optical parametric oscillators (OPOs).  In such degenerate processes, the
effective Hamiltonian has the form~\cite{Yuen, LAWu, SQ-book, SQ-30}
\begin{eqnarray}
  \hat{H}
  = \frac{i \hbar \, \chi^{(2)}}{2}( \alpha^\ast\hat{a}^2 - \alpha\hat{a}^{\dag 2}),
\label{eq:OPO_hamiltonian}
\end{eqnarray}
where $\alpha$ denotes the complex amplitude describing the pump field's strength and phase; while $\chi^{(2)}$ denotes the second order nonlinear susceptibility.  In
Eq.~(\ref{mastereq}), $\gamma$ is the system energy damping rate, and
$\bar{n}=[\text{exp}(\hbar \omega_0/k_B T) -1]^{-1}$ accounts for the average environmental photon occupation number at the optical driving frequency $\omega_0$, due to our experiment's effective thermal reservoir at tempera\-ture $T$.



The corresponding Wigner current of an ideal OPO system described by Eq.~(\ref{eq:OPO_hamiltonian})
driven at optical frequency $\omega_0$, denoted as $\vec{J}_{\text{sys}}$, is illustrated in
Fig.~\ref{fig:Stagnation_Point_Current_Schematic}~(b), and has the form
\begin{eqnarray}
  \vec{J}_{\text{sys}} &=&
    \chi^{(2)} |\alpha| \left(\begin{array}{c}
                              x \; W \cos\theta + \frac{p}{\omega_0}W \sin\theta \vspace{0.15cm}
                              \\
                              \omega_0  \; x \;  W \sin\theta -  \; p \; W \cos\theta
                            \end{array}\right);
 \label{eq:JsysGeneral} \\
                       &=& \xi 
                           \left( \begin{array}{c} \pm \frac{p}{\omega_0} W  \vspace{0.05cm}
      \\ \pm \omega_0 \; x \; W  \end{array}\right); \quad \text{for} \quad \theta = \pm \pi/2,  \;
\label{eq:Jsys}
\end{eqnarray}
with the squeezing parameter $\xi = \chi^{(2)} \alpha$.

Note, the Wigner current stagnation point at the origin of an OPO system has an orientation winding
number topolo\-gi\-cal charge of $\omega = -1$. This orientation winding number is
defined as ~\cite{Ole_PRL13}  
\begin{eqnarray}
  \omega({\cal L}(x,p)) \equiv \frac{1}{2\pi}\oint_{{\cal L}(x,p)} \,d\phi,
  \label{eq:WindingNumber}
\end{eqnarray}
and tracks the Wigner current's orientation along a simple closed-loop ${\cal L}$, i.e., around a single
stagnation point $(x,p)$, see Fig.~\ref{fig:Stagnation_Point_Current_Schematic}, and
Fig.~\ref{fig:Fig_State_Current} below.

In open systems, damping processes occur due to the coupling to the environment and drive the
initial wave package toward the origin.  Figure~\ref{fig:Stagnation_Point_Current_Schematic}~(c) illustrates this for a damped classical harmonic
oscillator. Note that classically the damping would concentrate the state ever more. In the quantum
case, however, such a concentration would violate Heisenberg's uncertainty principle. Instead, our
open dissipative system is well described by Eq.~(\ref{mastereq}).
The corresponding Wigner current, $\vec{J}_{\text{env}}$, has a dissipative part,
$\vec{J}_{\text{damp}}$, and a diffusive part, $\vec{J}_{\text{diff}}$, illustrated in Figs.~\ref{fig:Stagnation_Point_Current_Schematic}~(d)
and~\ref{fig:Stagnation_Point_Current_Schematic}~(e), respectively, and of the form~\cite{open}
\begin{eqnarray}
  \label{eq:Jenv}
  \vec{J}_{\text{env}} &=& -\frac{\gamma}{2} W \left(\begin{array}{c} x
                                                       \\
                                                       p
                                                     \end{array}\right) -\frac{\gamma}{2} \frac{\hbar}{\omega_0} (\overline{n}+\frac{1}{2}) \left( \begin{array}{c}  \partial_x W
                                                                                                                                                     \\  \partial_p W
                                                                                                                                                   \end{array}\right), \\
                        &\equiv& \vec{J}_{\text{damp}} + \vec{J}_{\text{diff}}.
\end{eqnarray}

So far, all investigations of
Wigner's current have been theo\-re\-tical due to the lack of experimental capability in capturing the
quantum dynamics in real time.  Nevertheless, for slowly evolving systems, in general, it is almost
always possible to identify and control an effective time~$\tau_{\text{eff}}$ by changing a coupling
constant in the Hamiltonian of the system; that is the approach we use.

\begin{figure}[t]
\centering
\includegraphics[width=8.4cm]{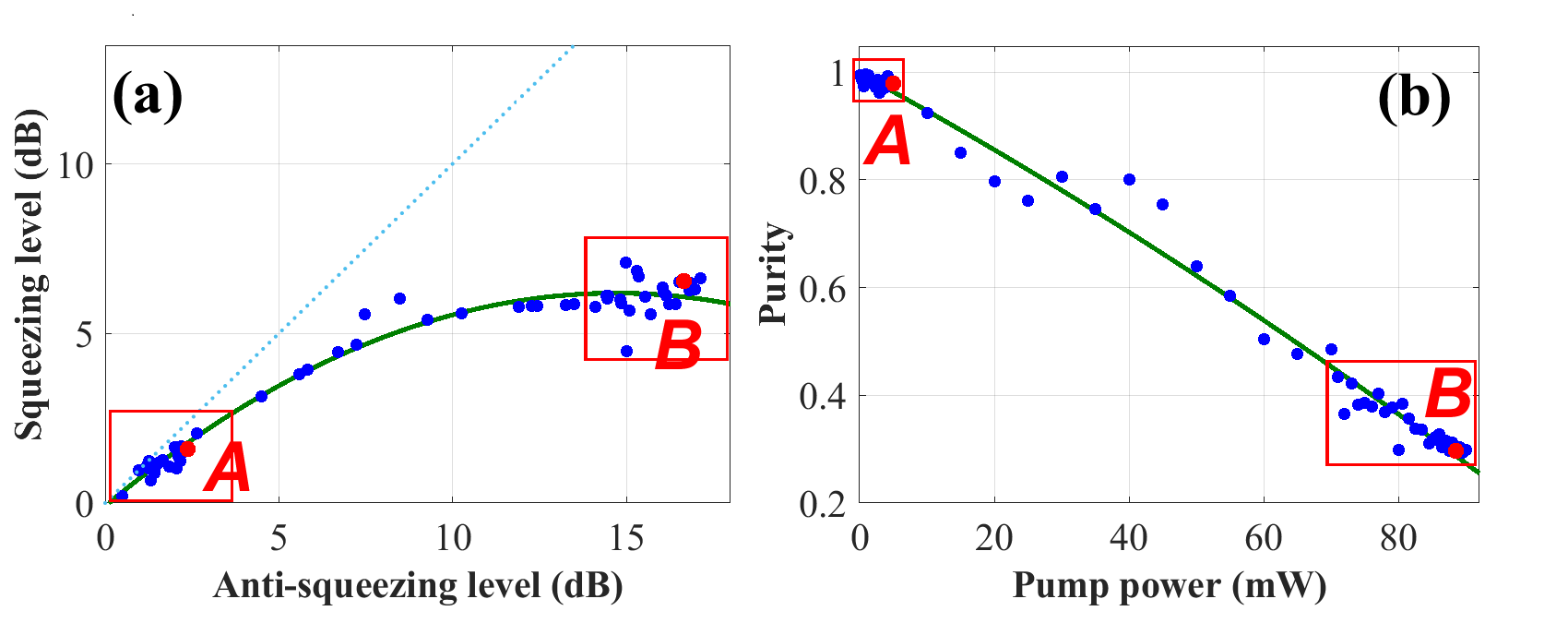}
\caption{(a) Measured quantum noise levels (in dB) of squeezing and anti-squeezing 
  quadratures at different pump powers. In the ideal case, squeezing and anti-squeezing levels should
    be equal (blue dashed line). Because of losses and phase noise, due to coupling to the environment, we instead observe degraded squeezing performance described by the solid green
    fit line. (b) The corres\-ponding purity,~$\text{tr}(\rho^2)$, of our squeezed states is determined using machine-learning enhanced quantum state tomography~\cite{QST-ML}. We highlight two regions with red boxes where the squeezing is `weak' and `strong'. Specific red markers $A$ and $B$ pick out data points generated for `low' and `high' pump power, which are referred to in Fig.~\ref{fig:Fig_State_Current}.}
     \label{fig:Purity_AntiSqueeze}
\end{figure}

\begin{figure*}[t!]
\centering
\includegraphics[width=16.8cm]{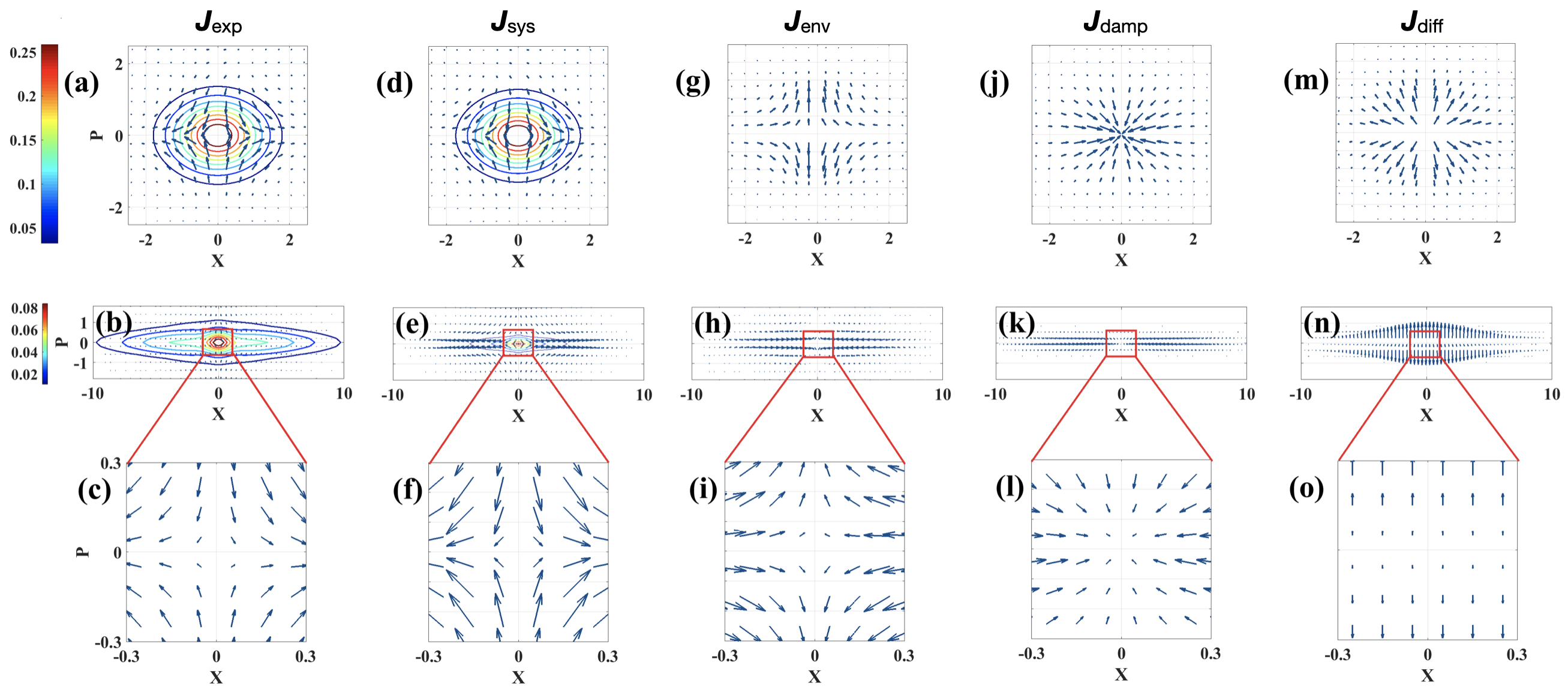}
\caption{Top and middle rows: snapshots of Wigner current distributions, displayed as blue arrows (in arbitrary units), overlaid over colored contours of the corresponding Wigner distributions of squeezed vacuum states. The top row represents a weakly squeezed state generated at an OPO pump power of $4.75$~mW and selected by marker $A$ in Fig.~\ref{fig:Purity_AntiSqueeze}. Similarly, the middle row represents a strongly squeezed state at $88$~mW pump power, selected by marker~$B$.
The Wigner current displays, circumscribed by red boxes in the middle row, are shown magnified in the bottom row. They illustrate how, even for strong coupling to the environment, the topological charge $\omega(0,0)$ of $\vec{J}_{sys}$ is preserved.
  First column (a-c) : Wigner current, $\vec{J}_{\text{exp}}$, reconstructed from experimental
  data. Second
  column (d-f): the ideal Wigner current, $\vec{J}_{\text{sys}}$, fitted to pure squeezed states. Third column (g-i): thermal
  contributions in the Wigner current $\vec{J}_{\text{env}} = \vec{J}_{\text{exp}} - \vec{J}_{\text{sys}}$, extracted from the
  experimental data (by subtraction of the Second column from the First column data). Fourth
  column (j-l): dissipative part of Wigner current, $\vec{J}_{\text{damp}}$ in Eq.~(\ref{eq:Jenv}). Fifth
  column (m-n): diffusive part of Wigner current, $\vec{J}_{\text{diff}}$ in Eq.~(\ref{eq:Jenv}). Note that adding $\vec{J}_{\text{damp}}$ and $\vec{J}_{\text{diff}}$ in Fourth and Fifth columns,  respectively, yields $\vec{J}_{\text{env}}$ shown in the Third column: $\vec{J}_{\text{env}} =\vec{J}_{\text{damp}} + \vec{J}_{\text{diff}}$.}
\label{fig:Fig_State_Current}
\end{figure*}

For an OPO Hamiltonian,~$\hat H$, as in Eq. (\ref{eq:OPO_hamiltonian}), the formal time evolution
operator has the form
\begin{eqnarray}
  \label{eq:U_evolution}
  \hat{U}(t) =\text{exp}[\frac{- i \hat H t}{\hbar}] = \text{exp}[\frac{\chi^{(2)}}{2}
  (\alpha^\ast\hat{a}^2 - \alpha\hat{a}^{\dag 2})], \;
\end{eqnarray}
of a squeezing transformation
$\hat{S} (\xi) = \text{exp}[\frac{1}{2} ( \xi^\ast \hat{a}^2 - \xi \hat{a}^{\dag 2})]$. This allows
us to connect our experimental parameters with the effective time parameter
$\tau_{\text{eff}} \propto |\xi| = \chi^{(2)} |\alpha|$ in conti\-nuity
equation~(\ref{eq:continuity}). In other words, varying the pump power amounts to a formal variation of
the evolution time and allows us to utilize the Wigner current~$\vec{J}$ of
Eq.~(\ref{eq:continuity}), also see\refAppendix{
}{}{}{sec:Appendix_ExpDensityMatrix}{}{.}\\

\noindent {\it Experiments.}\textemdash 
Here, we generate squeezed vacuum states in a bow-tie OPO cavity enclosing
a periodically poled nonlinear KTiOPO$_4$ (PPKTP) crystal with second order nonlinear susceptibility
$\chi^{(2)}$, operated below the lasing  threshold at the wavelength $1064$ nm~\cite{QST-ML}.
Such a bow-tie cavity can compensate the intra-cavity dispersion~\cite{bow-tie}. 

Our cavity has an optical path length of approximately $28.5$~cm and thus a free spectral range of $1.052$ GHz, with finesse of $19.8$ at $532$ nm and  of $33.4$ at $1064$ nm. The overall efficiency, defined as $1-L$ (the loss),  is $82.2 \pm 0.35\%$ and  phase noise is $34.50 \pm 1.26$ mrad.

By injecting the AC signal of our balanced homodyne detection, the spectrum analyzer records the
squeezing and anti-squeezing levels when scanning the phase of the local oscillator.  In
Fig.~\ref{fig:Purity_AntiSqueeze}~(a), we show the measured noise level curves for squeezing and anti-squeezing in decibel (dB) while the pump power increases from $0$ to $90$ mW. The magnitude of squeezing
and anti-squeezing levels are almost the same at low pump power levels, indicating
that the generated squeezed states are almost pure.  However, degradation arises due to the coupling
to the environment, giving roughly $7.08$ dB in sqeezing but $14.97$ dB in anti-squeezing at $76$ mW.
When the experiments are performed with pump power at $5$ mW, each photodiode of our balanced homodyne detector is illuminated with $15$ mW. 
The (electronic) dark noise level is $-87.69$ dB; while  the vacu\-um noise level is $-64.34$ dB, giving a shot noise level of $23.35$ dB. 
The measurement is done with the spectrum analyzer at $2.5$ MHz with $100001$ data points, $100$ kHz resolution bandwidth and $100$ Hz video bandwidth.
To make sure the measured squeezing level is not contaminated, our homodyne detectors are designed with a high common mode rejection ratio of more than $80$ dB~\cite{CMRR}.

We use machine learning to perform quantum state tomo\-graphy experimentally, after that a singular value decomposition of the  reconstructed density matrix gives three dominant terms
\begin{eqnarray}
\rho \approx \sigma_1\, \rho^{sq}+ c_1\, \rho^{sq}_{th} + d_1\, \rho_{th} .
\label{eq:mixedState}
\end{eqnarray}
Here, $\rho^{sq}$ denotes (pure) squeezed vacuum, $ \rho^{sq} =\hat{S}\rho_{0}\hat{S}^\dag$, with
$\rho_0$ the vacuum state and $S$ given by Eq.~(\ref{eq:U_evolution}). Due to coupling to the
environment, the two dominant admixtures are from thermal states $\rho_{th}$ and squeezed thermal
states $\rho^{sq}_{th} =\hat{S}\rho_{th}\hat{S}^\dag$~\cite{sq-thermal, sq-thermal-2, sq-thermal-3,
  sqthermal}. As shown in Fig.~\ref{fig:Purity_AntiSqueeze}~(b), the corresponding purity, $\text{tr}(\rho^2)$, of measured squeezed states decreases as the pump power increases.

To reconstruct Wigner's current we increase the pump power by $20$ discrete steps of $0.25$ mW each. This translates into effective time steps times~$\tau_j \propto |\alpha_j|$ which allows us to approximate $\partial W_{\text{exp}}(x,p) / \partial \tau_{\text{eff}}$ in the continuity equation~(\ref{eq:continuity}).

The corresponding Wigner current $\vec{J}_{\text{exp}}$ is then reconstructed using the parameters for the state given in Eq.~(\ref{eq:mixedState}), which in turn give us the effective weights with which we weigh the respective contributions from ideal
  system current~$\vec{J}_{\text{sys}}$ in Eq.~(\ref{eq:JsysGeneral}) and environment
  current~$\vec{J}_{\text{env}}$ in Eq.~(\ref{eq:Jenv}).  (For details see\refAppendix{}{}{}{sec:Appendix_ExpDensityMatrix}{}{).}
To our knowledge this shows for the first time a quantum system's experimentally reconstructed
Wigner current.

In Fig.~\ref{fig:Fig_State_Current}~(a-c), we observe that the experimentally reconstructed currents $\vec{J}_{\text{exp}}$ for squeezed vacuum
states follow hyperbolic curves aligned with the squeezed and anti-squeezed quadratures.
Compared to the pure system current $\vec{J}_{\text{sys}}$ predicted by theo\-ry in Eq.~(\ref{eq:Jsys}), shown in
Fig.~\ref{fig:Fig_State_Current}~\nolinebreak[4]{(d-f)},\linebreak[4] the experimentally observed current $\vec{J}_{\text{exp}}$,
displayed in Fig.~\ref{fig:Fig_State_Current}~\nolinebreak{(a-c)}, shows modifications due to decoherence
processes: expansions of the Wigner distributions and some distortions in the currents.
Their difference $\vec{J}_{\text{env}} = \vec{J}_{\text{exp}} - \vec{J}_{\text{sys}}$ is displayed
  in the Third column of Fig.~\ref{fig:Fig_State_Current}~(g-i). It graphically displays, to our
knowledge for the first time, how the system's current is locally counteracted by the currents due
to the coupling to the environment.
\\\\
\noindent{\it Low power, weak squeezing} ($P=4.75$~mW).\textemdash
To analyze and deepen our understanding of the roles played by the environment currents further, let us use Eq.~(\ref{eq:Jenv}) to decompose the Wigner distribution current
$\vec{J}_{\text{env}}$ into the dissipative $\vec{J}_{\text{damp}}$ and diffusive part
$\vec{J}_{\text{diff}}$, see the Fourth and Fifth columns of Fig.~\ref{fig:Fig_State_Current}.  Here, the
experimentally determined Wigner function $W_{\text{exp}}$ is used to generate the dissi\-pative and
diffu\-sive currents, $\vec{J}_{\text{damp}}$ and $\vec{J}_{\text{diff}}$, employing the two fitting
parameters, $\gamma=0.01$ and $\bar{n} = 0.1$.  These were extracted from the respective three
weights in Eq.~(\ref{eq:mixedState}) which change little as the pump power is increased to a low
power of $5$ mW, see Fig.~\ref{fig:Purity_AntiSqueeze}. This confirms that our system is stable
since losses due to the environment are fairly constant.
This squeezer system is exquisitely controllable and thus `clean' enough to allow us to emphasize what conceptual clarity and experimental resolving-power our toolbox of Wigner current measurements provides.

We find $ \vec{J}_{\text{diff}}$ in
Eq.~(\ref{eq:Jenv}) is dominated by the quantum vacuum contribution
$ \vec{J}_{\text{diff}} \approx -\frac{\gamma}{2} \frac{\hbar}{\omega_0} \frac{1}{2} (
  \partial_x W , \partial_p W )^\text{T} $, as sketched in Fig.~\ref{fig:Stagnation_Point_Current_Schematic}~(e).
We emphasize that, in order to make them clearly comparable with $\vec{J}_{\text{exp}}$ and
$\vec{J}_{\text{sys}}$ in Fig.~\ref{fig:Fig_State_Current}~(a, d), we had to increase the magnitudes of $\vec{J}_{\text{env}}$,
$\vec{J}_{\text{damp}}$ and $\vec{J}_{\text{diff}}$ in Fig.~\ref{fig:Fig_State_Current}~(g, j, m) by scale factors $375$,
$125$ and $125$, respectively. This demonstrates the impressive control over our system and
sensitivity to details of the dynamics and thus the strength of our approach 
(for more details see\refAppendix{
}{}{}{sec:Appendix_CurrentLowPower}{}{).}

\noindent{\it High power, strong squeezing} ($P=88$~mW).\textemdash
In our system, strong pumping, leading to strong squeezing simultaneously increases coupling
  to the environment and thus a pronounced reduction in the state's purity, see
  Fig.~\ref{fig:Purity_AntiSqueeze}~(b). The more squeezed the state, the less it shows
rotational symmetry in phase
space
, as can be seen clearly in Fig.~\ref{fig:Fig_State_Current}~(b).

  Yet, irrespective of the strength of the coupling to the environment, the environment's
  influence does not change the origin stagnation point's topological charge
  $\omega({\cal L}(0,0)) = -1$. This charge is topologically protected.

The protection can be understood as the OPO squeezer supplying the primary driving force, against this
  the environment reacts with dissipation and diffusion. Since the environment `reacts', it only
  `compensates' but cannot `overcome' the OPO's driving to impose a dynamic with entirely different
  topological characteristics. It thus reduces the effect of the OPO's driving without fundamentally
  altering the origin's charge of $\omega=-1$.

Note that in the second row of Fig.~\ref{fig:Fig_State_Current} the scale factor shown is $1$, indicating the low purity of the squeezed states.
We want to remark that  $\vec{J}_{\text{diff}}$ is state dependent [but $\vec{J}_{\text{damp}}$ is not, see Fig.~\ref{fig:Fig_State_Current}~(k, l) as well as 
  Eq.~(\ref{eq:Jenv})], this strongly modifies the 
  diffusive Wigner current, see Fig.~\ref{fig:Fig_State_Current}~(n, o). Together they form the overall
environmental Wigner current shown in  Fig.~\ref{fig:Fig_State_Current}~(h, i), which is
counteracting the system's squeezing action.
In terms of the underlying physical mechanisms we interpret our results as a {\it
    push-and-pull} scenario playing out in front of our eyes:

The squeezer-system current $\vec{J}_{\text{sys}}$ `pushes' the state along the $x$-axis whilst the environmental current $\vec{J}_{\text{env}}$ counteracts by `pulling backwards'. $\vec{J}_{\text{env}}$ , in turn, should be interpreted as the interplay between damping due to emission, see Fourth column of Fig.~\ref{fig:Fig_State_Current} [and also 
  Fig.~\ref{fig:Stagnation_Point_Current_Schematic}~(d)], and  quantum-diffusion due to absorption, see Fifth column of Fig.~\ref{fig:Fig_State_Current} [and Fig.~\ref{fig:Stagnation_Point_Current_Schematic}~(e)].
In the steady state these processes give rise to the Einstein coefficients. Here, for the first
time, we display their associated dynamical phase space signatures.\\

\noindent {\it Conclusion.}\textemdash 
With the help of machine learning-enhanced quantum state tomography, we experiment\-al\-ly
reconstruct the Wigner distribution and its phase space current of squeezed vacuum states through
the one-to-one mapping between the pump power and an effective time parameter. We find a Wigner
current stagnation point at the origin and confirmed its orientation winding number topological
charge as $\omega = -1$ and that it is topologically protected~\cite{Ole_PRL13}. The analysis of the Wigner current due to interactions with the thermal
environment reveals a push-and-pull between its damping and diffusive parts.

In addition to the squeezed states investigated here, our methodology can be readily applied to
other families of quantum states, such as single-photon states, `cat' states, and
quantum-optical engineering states~\cite{cat1, cat2, cat3, 2level}, for the  studies on their
evolution and interactions with outside systems.  Moreover, with recent theoretical developments in
systems described by anharmonic Hamiltonians~\cite{Skodje_PRA89, Oliva_Shear_19}, effectively
non-Hermitian parity-time symmetric Hamiltonians~\cite{PT} and even in discrete (spin)
systems~\cite{flow-sphere, sphere}, our experimental implementation promises to provide us with a
powerful dia\-gnostic toolbox.
The exprimental demonstration  allows us  to probe details of a quantum state's dynamics at levels
previously not accessible in various types of experiments~\cite{negativity, beyond-exp}, including in quantum
  optics~\cite{QST-Furusawa}, ultracold atoms~\cite{QST-atom1, QST-atom2},
  ions~\cite{QST-ion-1,QST-ion-2}, and superconducting devices~\cite{QST-SC}.

\section*{Acknowledgements}
This work is partially supported by the Ministry of Science and Technology of Taiwan
(Nos. 108-2923-M-007-001-MY3 and 110-2123-M-007-002), Office of Naval Research Global, US Army
Research Office (ARO), and the collaborative research program of the Institute for Cosmic Ray
Research (ICRR), the University of Tokyo.

\bibliography{ExpWigCurr_doi}

\clearpage

\setcounter{section}{0}
\renewcommand{\thesection}{Supplement~\arabic{section}}
\renewcommand{\thefigure}{S.~\arabic{figure}}
\setcounter{figure}{0}
\setcounter{equation}{0}
\renewcommand{\theequation}{S.\arabic{equation}}

\onecolumngrid
\vspace{\columnsep}

\begin{center}
  {   \large \bf -- Supplementary Materials -- \\ \vspace{0.25cm}

{Experimental Demonstration of Topological Charge Protection in Wigner Current}}
\\ \vspace{0.25cm}

{Yi-Ru Chen, Hsien-Yi Hsieh, Jingyu Ning, Hsun-Chung Wu, Hua Li Chen, You-Lin Chuang, Popo Yang, Ole
  Steuernagel, Chien-Ming Wu, and Ray-Kuang Lee}

 \end{center}



\section{Wigner Current Reconstruction\label{sec:Appendix_ExpDensityMatrix}}

How we reconstruct $\vec{J}_{\text{exp}}$:\\

Under the slowly varying pumping power condition, we need to find  $\delta J_{x}$ and $\delta J_{p}$. We do this by minimizing their 1-norm:

\begin{eqnarray}
    & \underset{\delta J_{x},\delta J_{p}}{\text{minimize}}
    & || \begin{bmatrix}
    \delta J_{x}\\
    \delta J_{p}
    
    \end{bmatrix}|| _{1},
\end{eqnarray}
subject to the continuity equation:
\begin{equation}
    \frac{\partial }{\partial t}W+\frac{\partial }{\partial x}(J^{\text{initial}}_{x}+\delta J_{x})+\frac{\partial }{\partial p}(J^{\text{initial}}_{p}+\delta J_{p})=0.
\end{equation}
Here, $J^{\text{initial}}_{x}$ and $J^{\text{initial}}_{p}$ are assigned as the initial guesses. 

Equivalently, we treat this as an optimization problem, taking the continuity equation as an equality constraint. Then, using machine learning-enhanced quantum state tomography~\cite{QST-ML}, we reconstruct the density matrix, as well as the correpsonding $W(x, p, \tau_{\text{eff}})$, from the experimental data at different effective times.

Explicitly, we treat $\delta J_{x}$ and $\delta J_{p}$ as the variables and solve the set of equations:
\begin{equation}
   D_{x}\,\delta J_{x}+D_{p}\, \delta J_{p} = \frac{W_{t}-W_{t+1}}{\Delta t}-D_{x}\,J_{x}^{\text{initial}}-D_{p}\, J_{p}^{\text{initial}}.
\label{Discrete}
\end{equation}
Here, the unknown variables are  $\delta J_{x} \in R^{n^{2}\times1}$, and $\delta J_{p} \in R^{n^{2}\times1}$, with $n$ being  the grid number in both $x$ and $p$ coordinates.
To perform the calculations on a discrete version of the continuity equation, we also have applied differential operators (forward finite difference), i.e., the differential matrixes $D_x$ and $D_p$ shown in Eq.~(\ref{Discrete}).
Finally, this $1$-norm  minimization problem can be transformed into a standard linear programming problem by using reformulation techniques~\cite{NoceWrig06}. 

\newpage
\setcounter{figure}{1}

 \section{Reconstructed Currents at Low Power:\\ High Resolving Power of our Approach\label{sec:Appendix_CurrentLowPower}}

\begin{figure*}[h]
\centering
\includegraphics[width=16.8cm]{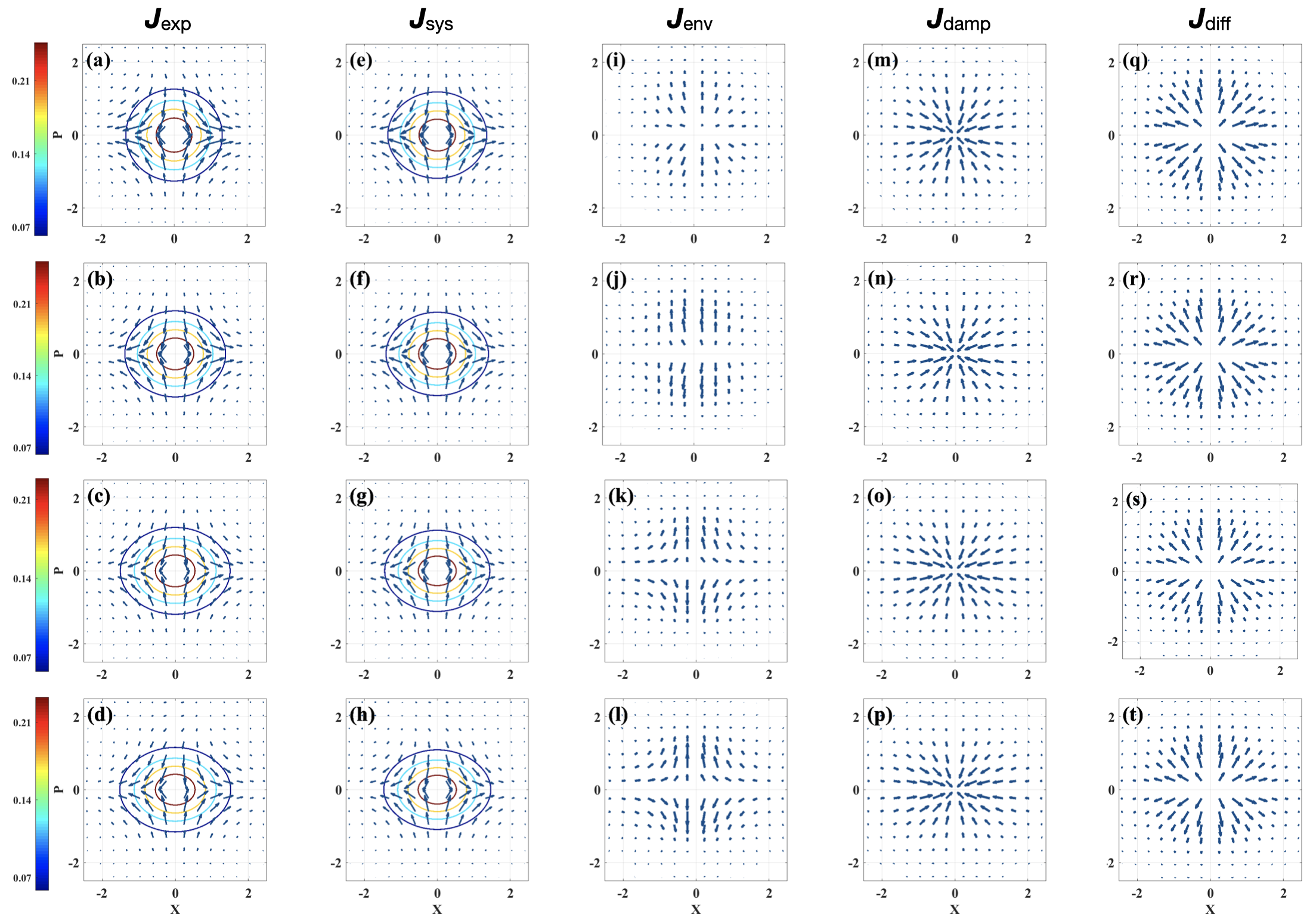}
\caption{Snapshots of Wigner current distributions (arrows, displayed in arbitrary units) of
  squeezed vacuum states at moderate squeezing, with the colored contours of corresponding Wigner
  distributions shown in the background. From Top to Bottom, the corresponding OPO pump powers are
  $0.25$, $1.75$, $3.25$, and $4.75$ mW, From Left to Right show the experimental Wigner current
  $\vec{J}_{\text{exp}}$, the ideal Wigner Current $\vec{J}_{\text{sys}}$, the environmental current
  $\vec{J}_{\text{env}} = \vec{J}_{\text{exp}} - \vec{J}_{\text{sys}}$, the damping current
  $\vec{J}_{\text{damp}}$, and the diffusive current $\vec{J}_{\text{diff}}$, respectively. Note
  that, we also have $\vec{J}_{\text{env}} = \vec{J}_{\text{damp}} + \vec{J}_{\text{diff}}$. }
\label{fig:State_and_Current__low_power}
\end{figure*}

For very moderate squeezing,
information about the degradation of the purity, $\text{tr}(\rho^2)$, of the quantum state in our
squeezing setup is shown in Fig.~\ref{fig:Purity_AntiSqueeze}~(b) (of the main text) as a function of the pump power. The purity of our
squeezed vacuum remains as high as $\approx 0.98$, even working at $5$ mW pump power. Yet, unavoidable
decoherence from the interaction with the environment is in evidence.
Its effects on the Wigner current are shown in Fig.~\ref{fig:State_and_Current__low_power},
which displays a series of experimentally determined snapshots of Wigner current
$\vec{J}_{\text{exp}}$ as the pump power is increased; the colored contours of corresponding Wigner
distributions are shown in the background.

Fig.~\ref{fig:State_and_Current__low_power} -- First column~(a-d): Wigner current, $\vec{J}_{\text{exp}}$, reconstructed from experimental
  data, at OPO pump powers $0.25$, $1.75$,  $3.25$,  and $4.75$ mW, respectively.  Second
  column~(e-h): the ideal Wigner current, $\vec{J}_{\text{sys}}$, fitted to pure squeezed states. Third column~(i-l): thermal
  contributions in the Wigner current $\vec{J}_{\text{env}} = \vec{J}_{\text{exp}} - \vec{J}_{\text{sys}}$, extracted from the
  experimental data (by subtraction of the Second column from the First column data). Fourth
  column~(m-p): dissipative part of Wigner current, $\vec{J}_{\text{damp}}$ in Eq.~(\ref{eq:Jenv}). Fifth
  column in Fig.~\ref{fig:State_and_Current__low_power}~(q-t): diffusion part of Wigner current, $\vec{J}_{\text{diff}}$ in Eq.~(\ref{eq:Jenv}). 
  (Note,
  adding $\vec{J}_{\text{damp}}$ and $\vec{J}_{\text{diff}}$ in the Fourth and Fifth columns,
  respectively, yields $\vec{J}_{\text{env}}$ shown in the Third column, i.e., $\vec{J}_{\text{env}} =\vec{J}_{\text{damp}} + \vec{J}_{\text{diff}}$.)

We emphasize that, in order to make them clearly comparable with $\vec{J}_{\text{exp}}$ and
$\vec{J}_{\text{sys}}$ in Fig.~\ref{fig:State_and_Current__low_power}~(Columns `1' and `2'),
  we had to increase the magnitudes of $\vec{J}_{\text{env}}$,
$\vec{J}_{\text{damp}}$ and $\vec{J}_{\text{diff}}$ in Fig.~\ref{fig:State_and_Current__low_power},~Columns `3', `4' and `5', by scale factors $375$,
$125$ and $125$, respectively. This demonstrates the impressive control over our system and
sensitivity to details of the dynamics and thus the strength of our approach 


\end{document}